\documentclass[traditabstract]{aa} 
\usepackage{longtable}
\usepackage{hyperref}
\usepackage{natbib}
\usepackage{graphicx}
\usepackage{txfonts}

\usepackage{color}
\usepackage{xcolor}

\def \kms{\ifmmode{~{\rm km\,s}^{-1}}\else{~km~s$^{-1}$}\fi}
\def \vhel{\ifmmode{V_{{\rm hel}}}\else{$V_{{\rm hel}}$}\fi}
\def \vsys{\ifmmode{V_{{\rm sys}}}\else{$V_{{\rm sys}}$}\fi}
\def \degree{\ifmmode{^{\circ}}\else{$^{\circ}$}\fi}

\def \myr{\ifmmode{{\rm\ M}_\odot{\rm\ yr}^{-1}}\else{${\rm\ M}_\odot$ 
yr$^{-1}$}\fi}

\def \mdot{\ifmmode{{\rm\dot{M}}}\else{${\rm\dot{M}}$}\fi}
\def \msun{\ifmmode{{\rm\ M}_\odot}\else{${\rm\ M}_\odot$}\fi}
\def \rsun{\ifmmode{{\rm\ R}_\odot}\else{${\rm\ R}_\odot$}\fi}
\newcommand{\HA}{H$\alpha$}

\newcommand{\OIII}{[O\,{\sc iii}]\ $\lambda$5007\,\AA}

\newcommand{\png}{PN~G283.7$-$05.1}

\begin{document}

\titlerunning{The post-CE central star of PN~G283.7$-$05.1}
   \title{The post-common-envelope binary central star of the planetary nebula  PN~G283.7$-$05.1\thanks{The radial velocity measurements and multi-band photometry of the central star of \png{} are available at the CDS via anonymous ftp to cdsarc.u-strasbg.fr (130.79.128.5)}} 

   \subtitle{A possible post-red-giant-branch planetary nebula central star}

   \author{D. Jones
          \inst{1,2}
          \and 
          H.~M.~J. Boffin \inst{3}
          \and
          J. Hibbert \inst{4,5}
          \and
          T. Steinmetz \inst{4,5}
          \and
          R. Wesson \inst{6}
          \and
          T.~C. Hillwig \inst{7}
          \and
          P. Sowicka \inst{8}
          \and
          \mbox{R.~L.~M. Corradi} \inst{9,1}
          \and
          J. Garc\'ia-Rojas \inst{1,2}
          \and
          \mbox{P. Rodr\'iguez-Gil}  \inst{1,2} 
          \and J. Munday \inst{10,1}
          }

    \institute{Instituto de Astrof\'isica de Canarias, E-38205 La Laguna, Tenerife, Spain\\
  \email{djones@iac.es}
              \and
              Departamento de Astrof\'isica, Universidad de La Laguna, E-38206 La Laguna, Tenerife, Spain
              \and
              European Southern Observatory, Karl-Schwarzschild-str. 2, D-85748 Garching, Germany
              \and
              Department of Physics and Astronomy, University of Sheffield, Sheffield, S3 7RH, UK
              \and
              Isaac Newton Group of Telescopes, Apartado de Correos 368, E-38700 Santa Cruz de La Palma, Spain
              \and
              Department of Physics and Astronomy, University College London, Gower Street, London WC1E 6BT, UK
              \and
              Department of Physics and Astronomy, Valparaiso University, Valparaiso, IN 46383, USA
              \and
              Nicolaus Copernicus Astronomical Center, Bartycka 18, PL-00-716 Warsaw, Poland
              \and
              GRANTECAN, Cuesta de San Jos\'e s/n, E-38712, Bre\~na Baja, La Palma, Spain
              \and
              Astrophysics Research Group, Faculty of Engineering and Physical Sciences, University of Surrey, Guildford, Surrey, GU2 7XH, United Kingdom
                           }

   \date{Received 26 June 2020 / accepted 31 August 2020}

 

 \abstract{
We present the discovery and characterisation of the post-common-envelope central star system in the planetary nebula \png{}.  Deep images taken as part of the POPIPlaN survey indicate that the nebula may possess a bipolar morphology similar to other post-common-envelope planetary nebulae.  Simultaneous light and radial velocity curve modelling reveals that the newly discovered binary system   comprises a highly irradiated M-type main-sequence star in a 5.9-hour orbit with a hot pre-white dwarf.  The nebular progenitor is found to have a particularly low mass of around 0.4~M$_\odot$, making \png{} one of only a handful of candidate planetary nebulae that is the product of a common-envelope event while still on the red giant branch.  In addition to its low mass, the model temperature, surface gravity, and luminosity are all found to be consistent with the observed stellar and nebular spectra through comparison with model atmospheres and photoionisation modelling.  However, the high temperature (T$_\mathrm{eff}\sim$95kK) and high luminosity of the central star of the nebula are not consistent with post-RGB evolutionary tracks.
}


   \maketitle
%

\section{Introduction}

It is now beyond doubt that central star binarity plays an important role in the formation and evolution of a significant fraction of planetary nebulae \citep[PNe;][]{soker97,jones17c,boffin19}.  However, the exact nature of that role, and its importance for the PN population as a whole, is still not well understood.  The strongest impact binary evolution can have on a resulting PN is via the so-called common-envelope (CE) phase, through which some 20\% or more PNe are believed to have been formed \citep[see e.g.][]{miszalski09a}.  Such an evolution has been shown to have a clear shaping effect on the resulting PN \citep{hillwig16}.  However, we still lack a statistically significant sample of these post-CE central stars for which the stellar and orbital parameters are well constrained enough to be able to effectively probe the properties of, and processes at work in, the CE phase.  Recent works have shown that detailed studies of post-CE central stars can provide critical information for the understanding of the CE phase, from the pre-CE mass transfer evolution \citep[e.g.][]{boffin12b,miszalski13b,jones15} through to the efficiency of the ejection \citep{iaconi19}.

\begin{figure*}[]
\centering
\includegraphics[width=\textwidth]{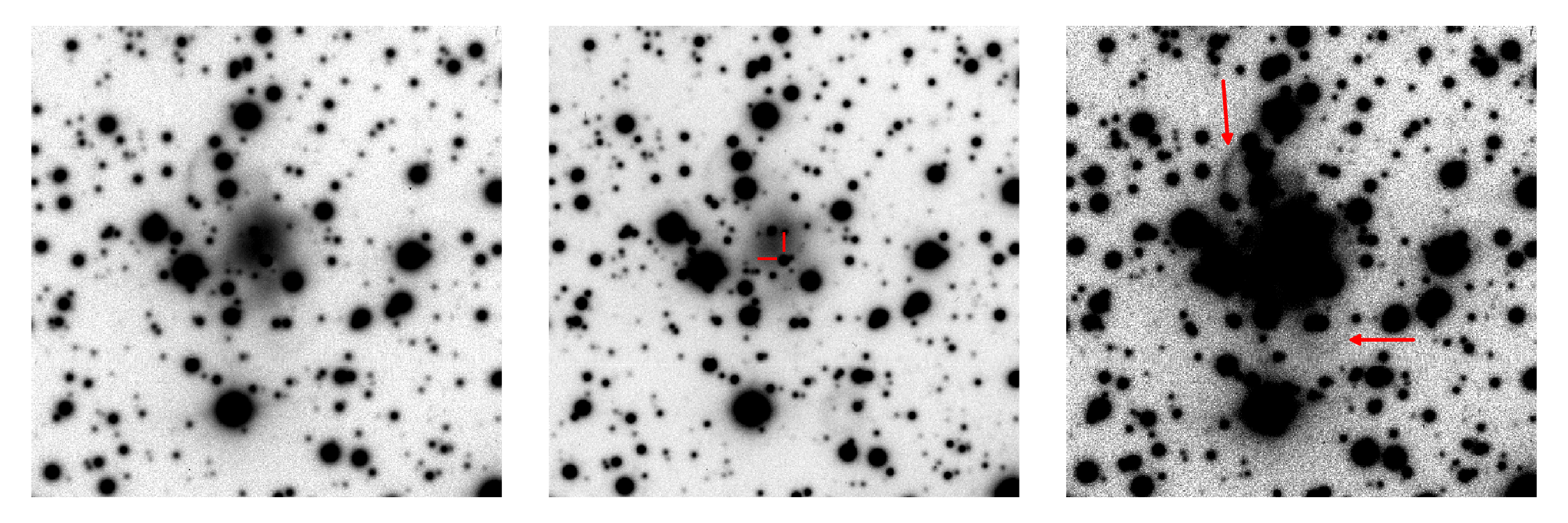}
\caption[]{POPIPlaN imagery of PN~G283.7$-$05.1 taken in the light of H$\alpha$+[N~\textsc{ii}] (left and right) and \OIII{} (centre).  The location of the central star is identified in the \OIII{} image, while on the higher contrast presentation of the H$\alpha$+[N~\textsc{ii}] image (right) the location of the `arcs' is indicated by the red arrows.  Each image measures 2\arcmin$\times$2\arcmin.  North is up, east is left.}
\label{fig:PNG283ims}
\end{figure*}

Here we present the latest results of an ongoing, concerted effort to reveal and characterise the binary nature of the central stars of PNe (CSPNe), in the form of the discovery and combined light and radial velocity curve modelling of the binary central star of \png{}. \png{} (PHR~J0958$-$6126; $\alpha$=09$^h$58$^m$32.3$^s$, $\delta$=$-$61\degr{}26\arcmin{}40\arcsec{}, J2000) was discovered by \citet[as part of the MASH project]{parker06}; it was classified as a `likely PN' and described as a `small, approximately circular, faint nebula'.  Further imagery was acquired as part of the POPIPlaN survey \citep{POPIPLAN} hinting at a more complex nebular structure than initially revealed by the shallower and lower resolution survey data used by the MASH project.  The POPIPlaN images (Figure \ref{fig:PNG283ims}) show the same roughly elliptical central nebula with an apparently off-centre central star, but also an arc of emission, visible in both \HA{} and [O~\textsc{iii}] images, located to the north-east (around 30\arcsec{} from the central star), as well as a possible counterpart in the south-west (around 22\arcsec{} from the central star).  The central region measures roughly 20\arcsec{}$\times$10\arcsec with a boxy appearance somewhat similar to that of Ou~5 \citep{corradi14}, possibly indicating that the nebula may actually be bipolar and that the arcs of emission represent more extended lobes.

This paper is organised as follows.  The observations and data reduction are presented in Section \ref{sec:obs}. The \textsc{phoebe}2 modelling of the central star system can be found in Section \ref{sec:phoebe};  an analysis of the nebular spectrum is presented in Section \ref{sec:neatalfamoc}.  Finally, the results are discussed in Section \ref{sec:disc}.

\section{Observations and data reduction}
\label{sec:obs}
\subsection{Photometry}
\label{sec:phot}
Between 22 and 23 March 2015 and then again between 2 and 6 March 2016, imaging observations of the central star system of PN~G283.7$-$05.1 were acquired using the EFOSC2 instrument \citep{EFOSC2a,EFOSC2b} mounted on the European Southern Observatory's 3.6 m New Technology Telescope (ESO-NTT).  The E2V CCD (pixel scale 0.24\arcsec{} pixel$^{-1}$) was employed along with the standard broadband filter set: B Bessel (\#639), V Bessel (\#641), R Bessel (\#642) and Gunn i (\#705).  Roughly $\sim$20 observations were made using each filter, with the exception of  B Bessel and R Bessel where nearly 100 were made.  The exact dates of each observation  can be found in the online data, while individual exposure times can be acquired (along with the unprocessed data) from the ESO archive\footnote{http://archive.eso.org/}.

The observations were debiased and flat-fielded using routines from the \textsc{AstroPy} affiliated package \textsc{ccdproc}.  Differential photometry of the central stars was then performed against field stars using the \textsc{sep} implementation of the \textsc{SExtractor} algorithms \citep{bertin96,sep} before being placed on an apparent magnitude scale using observations of standard stars taken during the course of the observations.  A circular aperture of 1'' was used for the photometry in order to avoid contamination from a field star roughly 2'' to the west of the CSPN.  We note that the nebula itself is too faint to significantly contaminate the photometry, even in the broadband filters used.  The resulting photometric measurements are available in the online data.

\begin{figure*}[]
\centering
\includegraphics[width=\textwidth]{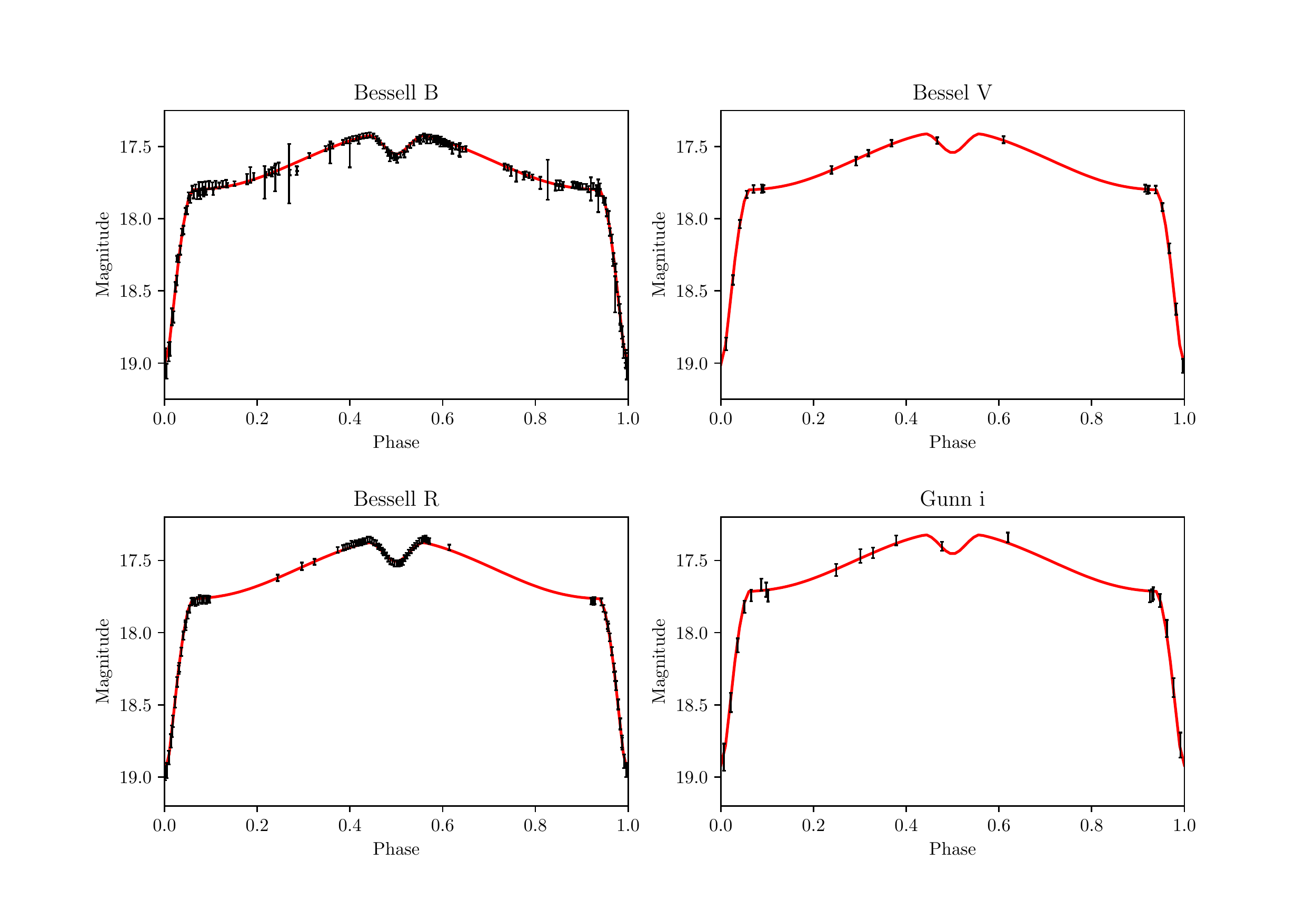}
\caption[]{Phase-folded photometry of the central star system of \png{} overlaid on the best-fitting \textsc{phoebe}2 model light curves. 
}
\label{fig:PNG283phot}
\end{figure*}

The light curves were searched for periodicities using the \textsc{period} package of the \textsc{starlink} software suite.  The determined ephemeris is

\begin{equation}
\mathrm{HJD}_\mathrm{eclipse}=2\,457\,449.883246(1) +  0.245845(3) E
\end{equation}
for the Heliocentric Julian Date of the mid-point of the deepest (primary) eclipse ($\mathrm{HJD}_\mathrm{eclipse}$), where $E$ is an integer representing the number of orbits since the eclipse time quoted in the ephemeris.  The phase-folded light curves for each filter are shown in Figure \ref{fig:PNG283phot}.

The primary eclipse is relatively well observed in all bands, with depth decreasing with the redness of the passband, entirely consistent with the primary being much hotter than the secondary and thus contributing more in the bluer bands.  The secondary eclipse is also clearly detected in both bands (B and R) for which observations were taken at the appropriate phase.  These eclipses are observed superimposed upon a roughly sinusoidal modulation with minimum at primary eclipse and maximum at secondary eclipse; this  light curve morphology is typical of the secondary being irradiated by the primary, again indicating that the primary must be significantly hotter than the secondary.

\subsection{Radial velocity monitoring}
\label{sec:spec}

Spectroscopic monitoring of the central star system of PN~G283.7$-$05.1 was performed on 5--9 January 2016 using the FOcal Reducer/low dispersion Spectrograph \citep[FORS2;][]{FORS} instrument mounted on the Unit Telescope 1 (UT1 or Antu) of the European Southern Observatory's 8.2m Very Large Telescope (ESO-VLT).  A 0.7\arcsec{}  slitwidth was employed along with the 1400V grism (approximate wavelength coverage, 4600\AA{}$\leq\lambda\leq$5800\AA{}, with a resolution of R$\sim$3000) and the blue-sensitive E2V detector.  Exposure times were 900 s; the exact dates of each individual observation can be found in Table \ref{tab:PNG283RV}.  Basic reduction (debiasing, flat-fielding, wavelength calibration) was performed using the standard FORS2 pipeline before sky-subtraction and optimal extraction of target spectra by standard \textsc{starlink} routines \citep{figaro}.  The resulting spectra were then corrected for heliocentric velocity via cross-correlation of the [O~\textsc{iii}] nebular lines at 4958.911\AA{} and 5006.843\AA{}, placing all spectra at the rest velocity of the host nebula.

\begin{figure}[]
\centering
\includegraphics[width=\columnwidth]{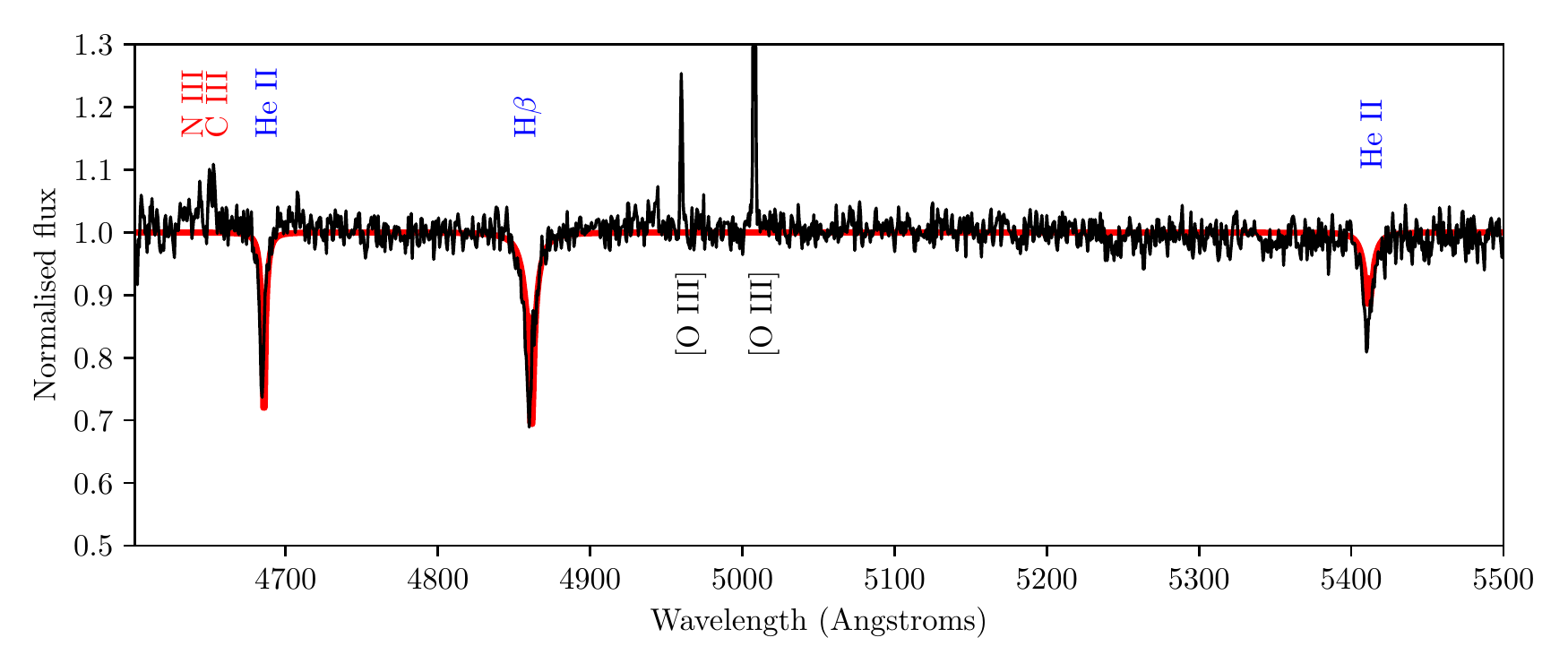}
\caption[]{Example spectrum of the central star of PN~G283.7$-$05.1 (in black), taken at an orbital phase of roughly 0.26.  The spectrum highlights the prominent absorption spectrum of the primary as well as the extremely weak nature of the irradiated emission line complex.  A TMAP synthetic spectrum with T$_\mathrm{eff}$ and surface gravity roughly consistent with the \textsc{phoebe}2 model is underlaid in red.}
\label{fig:png283spec}
\end{figure}

\begin{figure}[]
\centering
\includegraphics[width=\columnwidth]{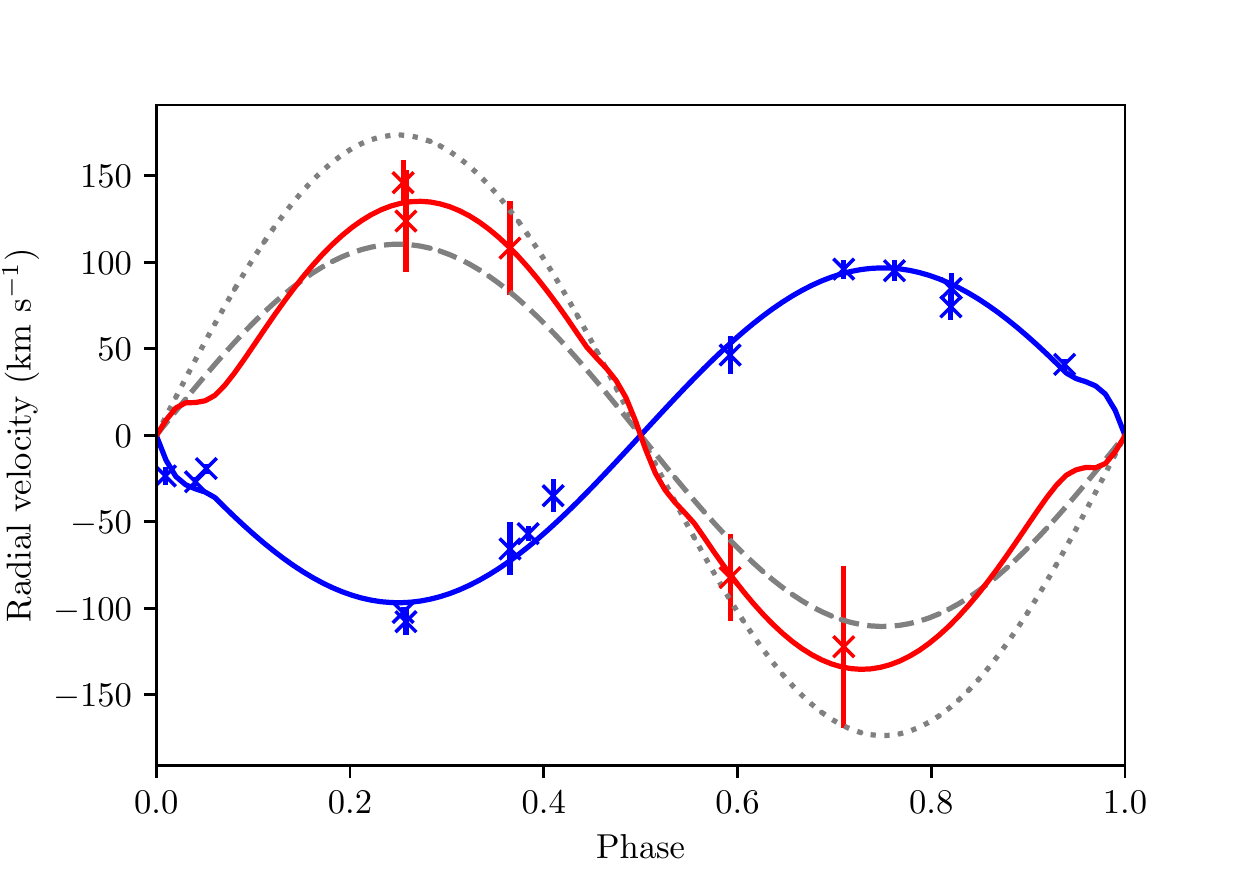}
\caption[]{Phase-folded RVs of the hot (blue) and cool (red) components of the central star system of \png{}, overlaid on the best-fitting \textsc{phoebe}2 model RV curves.  The centre-of-mass RV curve of the secondary is shown as the dotted grey line, while the RV of its innermost point (i.e.\ the most highly irradiated part of the stellar surface) is shown as the dashed grey line.}
\label{fig:png283rv}
\end{figure}

The spectrum of the central star of PN~G283.7$-$05.1 shows pronounced He~\textsc{ii} absorption lines at both 4685.61\AA{} and 5411.52\AA{} (see Figure \ref{fig:png283spec}).  However, the emission line complex C~\textsc{iii} and N~\textsc{iii} at 4630--4650\AA{}, associated with the irradiation of the secondary by the hot primary \citep{miszalski11b}, is only very weakly detected (or even undetected) at most phases (with the exception of observations taken between phases $\sim$0.3 and 0.7 when the irradiated face of the companion is most prominent).  Cross-correlation with a template spectrum containing the irradiated emission complex (N~\textsc{iii} 4643\AA{} and 4640\AA{}; C~\textsc{iii} 4647\AA{}, 4650\AA{}, and 4651\AA{}) was used to derive the radial velocities (RVs) of the companion, while a template with the prominent absorption features shown in Figure~\ref{fig:png283spec} (H$\beta$;  He~\textsc{ii} 4686\AA{} and 5412\AA{}) was used for the RVs of the primary.  In both cases, the uncertainties were estimated by fitting a bisector to the strongest peak of the cross-correlation function.  The resulting RV measurements are presented in Table \ref{tab:PNG283RV}, while the data is shown folded on the ephemeris derived from the more comprehensive photometric dataset (described in section \ref{sec:phot}) in Figure \ref{fig:png283rv}.  

Sine curves fitted to the RVs are in good agreement with the photometric ephemeris, with both components showing velocity zeros during eclipse and maxima or minima at quadrature.  The primary RV curve is well fit by a sine of amplitude K$_1\sim97$\kms{}.  The RV curve of the secondary is more sparsely sampled, but fitting results in an amplitude K$_2\sim140$\kms{}. However, the true centre-of-mass amplitude is likely much greater as the emission lines used to measure the radial velocity of the secondary are only produced in the irradiated hemisphere \citep{exter03,miszalski11a}.  Together, the RV curves imply a mass ratio, $q= \frac{m_2}{m_1} \leq0.7$.

\begin{table}
\caption{Radial velocity measurements for the two components of the central star of PNG283.7$-$05.1. The irradiated emission line complex was not detected in more
than half of the spectra, and thus the radial velocity of the cool component was not measured.}             
\label{tab:PNG283RV} 
\centering
\begin{tabular}{lrlrl}
\hline\hline
HJD & \multicolumn{2}{c}{Hot component} & \multicolumn{2}{c}{Cool component} \\
& \multicolumn{2}{c}{(km~s$^{-1}$)} & \multicolumn{2}{c}{(km~s$^{-1}$)} \\
\hline
2457393.78640 & 74.3 & $\pm$7.4 &\multicolumn{2}{c}{-}\\
2457393.81527 & 41.0 & $\pm$3.1 &\multicolumn{2}{c}{-}\\
2457394.66885 & $-$34.9 & $\pm$9.6 &\multicolumn{2}{c}{-}\\
2457394.71376 & 46.4 & $\pm$11.2 & $-$82.3 & $\pm$25.3 \\
2457394.76988 & 84.9 & $\pm$9.2 &\multicolumn{2}{c}{-}\\
2457394.81629 & $-$23.5 & $\pm$5.1 &\multicolumn{2}{c}{-}\\
2457395.64128 & $-$65.7 & $\pm$15.3 & 108.2 & $\pm$27.4 \\
2457395.72598 & 96.1 & $\pm$5.5 & $-$122.2 & $\pm$46.9 \\
2457395.81005 & $-$19.2 & $\pm$2.8 &\multicolumn{2}{c}{-}\\
2457395.86071 & $-$107.7 & $\pm$7.7 & 123.8 & $\pm$29.6 \\
2457396.72222 & 95.2 & $\pm$6.1 &\multicolumn{2}{c}{-}\\
2457396.79068 & $-$26.9 & $\pm$2.5 &\multicolumn{2}{c}{-}\\
2457396.84338 & $-$102.4 & $\pm$3.1 & 146.0 & $\pm$13.0 \\
2457396.87510 & $-$56.9 & $\pm$4.4 &\multicolumn{2}{c}{-}\\
\hline
\end{tabular}
\end{table}

\subsection{Nebular spectroscopy}
\label{subsec:nebspec}

On the night of 9 December 2015, the nebula of \png{} was observed using the FORS2 instrument employing the red-sensitive MIT/LL CCD mosaic and a 0.7\arcsec{}  slitwidth  at a position angle of $-20$\degr{} (spatial scale of 0.25\arcsec{} pixel$^{-1}$).  A 2500s exposure was acquired using the GRIS\_1200B grism (resolution $\sim$1\AA{} across the wavelength range 3600$\lesssim\lambda\lesssim$5000\AA{}), and a further 120s exposure using the GRIS\_600RI grism and the GG435 order blocking filter (resolution $\sim$3\AA{} across the wavelength range 5100$\lesssim\lambda\lesssim$8300\AA{}).  The raw data were bias subtracted, cleaned of cosmic ray hits, wavelength- and flux-calibrated (via calibrations and standard star observations taken as part of the standard ESO calibration plan) using standard \textsc{starlink} routines \citep{figaro,ccdpack}.  The reduced two-dimensional frames were then extracted to one-dimensional spectra using ESO-\textsc{midas}, avoiding the region containing the central star in order to isolate the nebular spectrum \citep[as in][]{wesson18}.

About 15 emission lines are detected in the nebular spectrum, including several hydrogen Balmer lines, and a few of the brightest lines of He~{\sc i}, He~{\sc ii}, [O~\textsc{iii}], [Ar~{\sc iii}], and [Ne~{\sc iii}] (see Table \ref{linelist}).  The extinction could be estimated from the Balmer line ratio of H$\alpha$, H$\beta$, and H$\gamma$ to be $\mathrm{c}(\mathrm{H}\beta$)=0.4$\pm$0.2, corresponding to a visual extinction, $A_V$, of approximately 0.77 
magnitudes.

\section{Simultaneous modelling of the light curves and RV  curves}
\label{sec:phoebe}

In order to derive the system parameters of the binary central star of \png{}, the light and radial velocity curves presented in Sections \ref{sec:phot} and \ref{sec:spec} were modelled simultaneously using the version 2.2 release of the \textsc{phoebe2} code \citep{phoebe2_2}.  

The hot primary was modelled as a blackbody with linear limb darkening (LD), where the intensity of a given stellar surface element ($I_\mu$) is taken to vary as a function of the emergent angle ($\theta$, with $\mu$ defined as the cosine of this angle) in the form $I_\mu/I_0=1-x(1-\mu)$, the coefficient of which ($x$) was considered a free parameter.  The secondary was modelled using Castelli \& Kurucz model atmospheres \citep{ck2004}, with interpolated LD as implemented in the \textsc{phoebe}2 code \citep{phoebe}.  The stellar masses, temperatures, and radii, as well as the binary inclination and secondary albedo, were considered free parameters.  The systemic velocity of the binary was fixed to the nebular velocity (i.e. zero on the scale of Figure~\ref{fig:png283rv}). As mentioned in Section \ref{sec:spec}, the RVs of the secondary do not reflect its centre-of-mass RV but rather that of the irradiated region in which the emission lines are produced.  By default, \textsc{phoebe}2 determines the flux-weighted (i.e.\ centre-of-light) RVs of the model binary components.  As the flux-weighted RVs are derived after the treatment of irradiation, this technique is essentially the same as the K-correction method applied to, for example, low-mass X-ray binaries \citep{munozdarias05} and cataclysmic variables \citep{thoroughgood05}.  Without the implementation of irradiated atmosphere models (which would predict the exact zones from which the irradiated emission lines originate), no further improvement is possible on the modelling approach \citep{horvat19}.  However, the possible implications of this approach on our conclusions will be explored in full.  

Following an initial manual exploration of the parameter space, fitting was performed via a Markov chain Monte Carlo (MCMC) method as outlined in \citet{jones19}. The final model light curves and RV curves are shown overlaid on the data in Figures \ref{fig:PNG283phot} and \ref{fig:png283rv}, respectively.  The model provides a good fit to all the data at all phases. The RV residuals are found to be of order one uncertainty or less for both components at all phases.  We note that the model RV curves for both components show appreciable Rossiter-McLaughlin effects as a result of the eclipsing nature of the system.  The RV curve of the irradiated companion shows a similar effect away from eclipse, due the transition of its photocentre from the irradiated hemisphere (which is disappearing from view at these phases) to the now-visible non-irradiated face.

\begin{figure*}[]
\centering
\includegraphics[width=\textwidth]{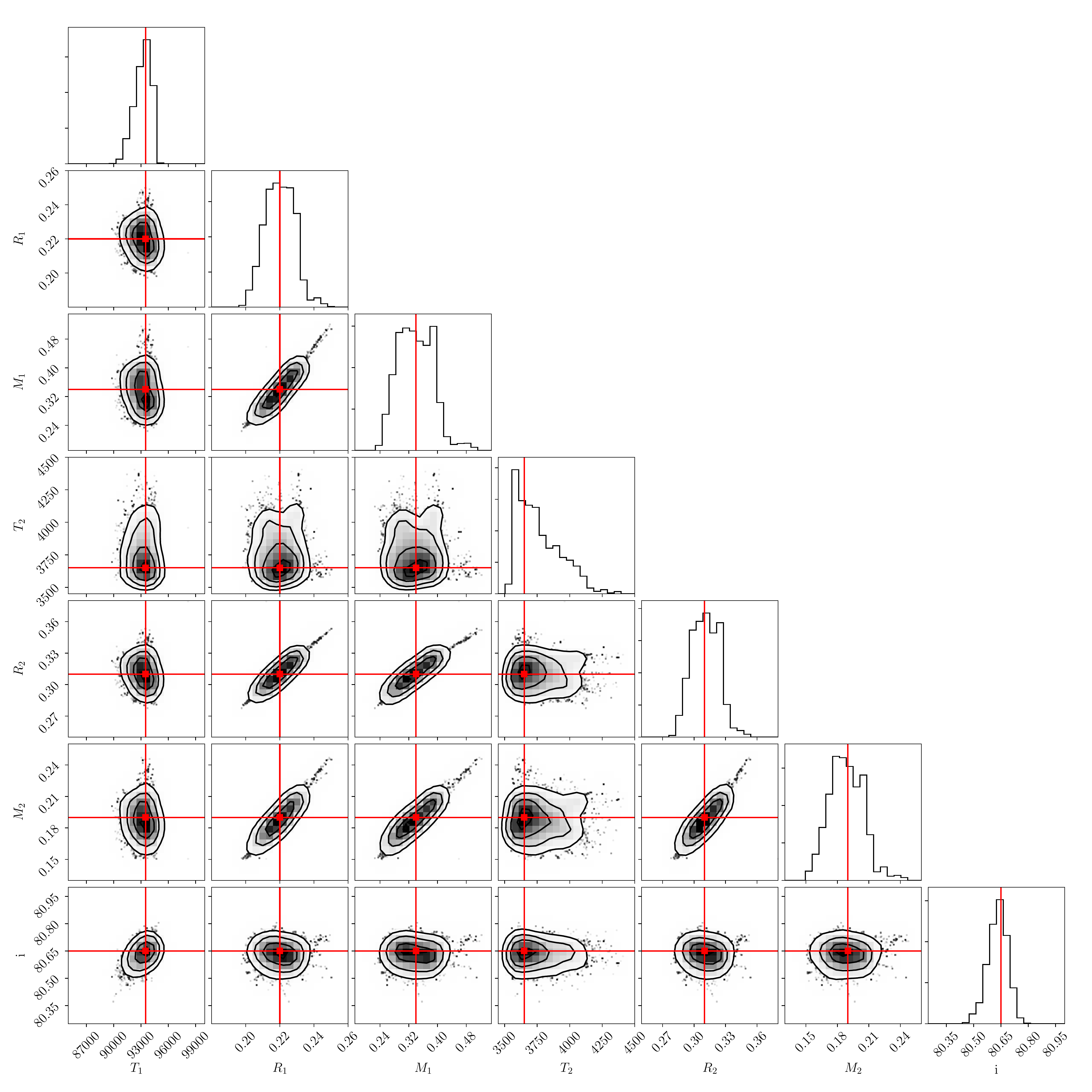}
\caption[]{Corner plot \citep[made using \textsc{corner};][]{corner} of the \textsc{phoebe}2 MCMC posteriors for the stellar temperatures ($T_1$, $T_2$), radii ($R_1$, $R_2$), and masses ($M_1$, $M_2$), as well as the binary orbital inclination ($i$).}
\label{fig:PNG283mcmc}
\end{figure*}

The fit to the photometric is good except in regions where the observations themselves display  a large intrinsic scatter\footnote{This scatter is likely due to underestimation of the photometric uncertainties, which are purely statistical and include no contribution from nebular contamination.} such as the B-band primary eclipse (where the deviation between observations and model are 3$\sigma$ at most but generally much lower,  particularly when the photometric data are binned).  The final model parameters and their uncertainties (from sampling of the MCMC posteriors, as presented in Figure \ref{fig:PNG283mcmc}) are shown in Table \ref{tab:CSparams}.  

\begin{table}
\centering
\caption{Parameters of the central stars of \png{} as determined by the \textsc{phoebe}2 modelling. }              
\label{tab:CSparams}      
\centering                                      
\centering
\begin{tabular}{lrlrl}
\hline\hline
& \multicolumn{2}{c}{Hot component} & \multicolumn{2}{c}{Cool component} \\
\hline
Mass (M$_\odot$) & 0.34&$\pm$0.05 & 0.19 &$\pm$0.02\\
Radius (R$_\odot$) & 0.22&$\pm$0.01 & 0.31 &$\pm$0.01\\
T$_\mathrm{eff}$ (kK) & 93.5&$\pm$1.5\tablefootmark{a} & 3.6&$\pm$0.2\\
Albedo & \multicolumn{2}{c}{1.0 (fixed)} & 1.0 &$\pm$0.01\\
Linear LD coeff.\ & 0.03&$\pm$0.03 & \multicolumn{2}{c}{interpolated\tablefootmark{b}}\\
\hline
Orbital period (days) && 0.245845&$\pm$0.000003 &\\
Orbital inclination & \multicolumn{2}{r}{80.65$^\circ$}&\multicolumn{2}{l}{$\pm0.1^\circ$}\\
\hline
\end{tabular}
\tablefoot{
\tablefoottext{a}{Based on the implied luminosity, the uncertainty on the primary temperature is likely underestimated.  See text for discussion.}
\tablefoottext{b}{Interpolated limb darkening, based on the \citet{ck2004} atmospheres as implemented in \textsc{phoebe}2 \citep{phoebe}, was used rather than a linear limb-darkening law.}
}
\end{table}

The model secondary radius is roughly 50\% larger than expected for its mass \citep[see e.g.][]{parsons18},   a finding that is not uncommon amongst young post-CE binaries, with most having been found to present much more inflated secondary radii, probably as a result of mass transfer \citep{jones15}.  The secondary temperature is also a little greater than would be expected for an isolated field star of the same mass and radius; however, given the high levels of irradiation as evident from the shape of the binary light curve, the model temperature seems reasonable.

The model primary mass is found to be particularly low, well-below the transition between post-RGB and post-AGB core masses even when accounting for the uncertainties.  However, its relatively high temperature and low surface gravity (log g$\sim$5.3) are not consistent with post-RGB evolutionary tracks \citep{hall13,hillwig17}.  The primary mass is principally constrained by the measured secondary RVs which, as highlighted above, do not directly reflect the centre-of-mass RV of the star.  If the region from which the irradiated emission lines originate is closer to the first Lagrange point than the B-band photocentre of the model, then the model primary mass will inevitably be underestimated.  To explore this possibility, a second series of MCMC chains were run, this time taking the secondary RV to be representative of only the point closest to the hot component (i.e.\ maximising the primary mass while still fitting the other observables\footnote{It is important to repeat the fitting rather than simply adjusting the primary mass while maintaining all other parameters the same; changing the primary mass also alters the orbital separation, and thus means other parameters must be altered in order to fit the observed light curve (namely the stellar radii).}).  In this case, the model primary mass increases from 0.34$\pm$0.05~M$_\odot$ to 0.42$\pm$0.05~M$_\odot$, while its radius hardly changes to 0.23$\pm$0.03~M$_\odot$.  As such, although the model primary mass quoted in Table \ref{tab:CSparams} may be underestimated, it is unlikely to be greater than $\sim$0.47~M$_\odot$, and therefore unlikely to be consistent with any post-RGB evolutionary tracks\footnote{Interestingly, the model luminosity and temperature are consistent with the post-AGB tracks of \citet{MillerBertolami2016} for a remnant mass of $\sim$0.54~M$_\odot$ and an age of $\sim10,000$ years.}.   The mass, temperature, and radius of the secondary vary by approximately one uncertainty with respect to the centre-of-light model, meaning that the conclusions regarding its parameters do not change.

The luminosity of the model primary component (log $L/L_\odot \sim3.5$) is extremely high for a post-RGB star, given that a typical maximum luminosity at the tip of the RGB is log $L/L_\odot \sim3.4$ \citep{salaris97}.  However, this maximum value is within two uncertainties of the luminosity of the model central star, implying that the discrepancy may not be irreconcilable. Furthermore, the luminosity is strongly dependent on the effective temperature ($\propto \mathrm{T}_\mathrm{eff}^4$), and the uncertainty on the primary temperature is almost certainly underestimated.  The model primary temperature is strongly dependent on two factors: the amplitude of the observed irradiation effect and the depth of the primary eclipse.  We have already highlighted that the treatment of irradiation can introduce additional uncertainties in the context of the observed secondary RVs, but it can also contribute here to the irradiation effect amplitude. As an irradiated star does not   simply display the spectrum of a hotter star, but displays emission lines that can contribute unequally between filter bands, the standard treatment of irradiation can cause problems in replicating the observed increase in systemic brightness across these bands. The depth of the primary eclipse should be a more robust estimate of the primary temperature, but again there are additional sources of uncertainty that are not accounted for in the values presented in Table \ref{tab:CSparams}.  The primary was modelled as a blackbody; any significant deviations from this in the true spectrum could have an impact.  Furthermore, the contribution of the secondary (to both the primary eclipse depth and amplitude of the irradiation effect) may be skewed by the use of the \citet{ck2004} model atmospheres, which only go as low as a $\mathrm{T_{eff}}$ of 3500 K.  If the true temperature of the secondary were lower than this limit then the observed amplitude of primary eclipse and irradiation effect could feasibly be reached with a slightly reduced primary temperature (implying a reduced luminosity).  Importantly, however, even allowing for an increased uncertainty on the primary temperature, the model is still not consistent with post-RGB tracks \citep[which would require much lower temperatures, in the range 60--70kK, for such a mass and radius;][]{hall13}.

As a consistency check of the model, synthetic photometry of the system was calculated using \textsc{synphot} \citep{synphot}. The two components were assumed to be blackbodies and reddened following the extinction law of \citet{cardelli89} with an $R_V$=3.1 (just as assumed in the \textsc{phoebe}2 modelling).  The observed B, V, R, and i magnitudes are then well reproduced by the model parameters, with a visual extinction $A_V$$\sim$1 \citep[roughly consistent with that measured from the nebular spectrum, as well as the field extinction of $A_V$=1.1 determined from Sloan Digital Sky Survey spectra by][]{schafly11} at a distance of approximately 9.5~kpc.  This distance seems extreme for the relatively low extinction, but the Galactic longitude of \png{} places it in a line of sight between the Sagittarius and Perseus arms of the Galaxy, for which the extinction may well be minimal.  Furthermore, the measured angular size of the nebula is not unreasonable at that distance, implying a physical size of roughly 1~pc.

The Gaia parallax of the central star of \png{} is 0.2184$\pm$0.0955 mas which, although rather imprecise, still permits the distance to be estimated. \citet{bailerjones18} derive a most probable distance of 4.8~kpc, with a confidence interval ranging from $\sim$2.5--8.8~kpc.  While the most probable value is certainly at odds with the distance derived by the \textsc{phoebe}2 modelling, once the uncertainties on the stellar parameters (equating to approximately $\pm$0.5~kpc in terms of distance) are considered, a borderline agreement is found.  Again, if an increased uncertainty on the primary temperature is accounted for, the agreement improves.

As a further consistency check, we compared synthetic spectra produced with the T\"ubingen Model Atmosphere Package \citep[TMAP;][]{rauch03,werner03} with the observed spectrum shown in Figure~\ref{fig:png283spec}.  Assuming a contribution of roughly 0.16 magnitudes at the observed phase of 0.26 (derived from the change in B-band brightness between this phase and the egress of primary eclipse), a reasonable fit (shown in red in Figure \ref{fig:png283spec}) is found for an H+He white dwarf\footnote{The model abundances are H:He=7:3, interestingly in line with the best-fitting post-AGB track of \citet{MillerBertolami2016} which has H:He=6.7:2.9, albeit for a higher mass than determined by the \textsc{phoebe}2 model.} T$_\mathrm{eff}$=90kK and log g=5.25 (cf.\ T$_\mathrm{eff}\sim$93kK and log g$\sim$5.3, as determined by the \textsc{phoebe}2 simultaneous light curve and RV curve modelling).  As such, the observed spectrum is found to be consistent with the modelled primary parameters.

\begin{table*}
 \caption{Observed and dereddened line fluxes relative to F(H$\beta$)=100 measured in the FORS2 spectrum of \png.}
 \label{linelist}
 \begin{tabular}{llrlrlllllll}
 \hline
 $ \lambda_{\rm obs} $ & $\lambda_{\rm rest}$ & \multicolumn{2}{c}{$F \left( \lambda \right) $} & \multicolumn{2}{c}{$I \left( \lambda \right) $} & Ion & Multiplet & Lower term & Upper term & g$_1$ & g$_2$ \\
 \hline
  3871.20 &   3868.75 &   40.9 & $\pm$   2.3 &  53.5 & $^{  +5.6 }_{  -6.2 }$ & [Ne~{\sc iii}]   & F1         & 2p4 3P     & 2p4 1D     &          5 &        5    \\
  3891.12 &   3888.65 &   17.7 & $\pm$   1.5 &  21.8 & $^{  +2.0 }_{  -2.2 }$ & He~{\sc i}       & V2         & 2s 3S      & 3p 3P*     &          3 &        9    \\
        * &   3889.05 &  *     &             & *     &                        & H~{\sc i}        & H8         & 2p+ 2P*    & 8d+ 2D     &          8 &        *    \\
  3969.95 &   3967.46 &   12.2 & $\pm$   1.0 &  14.3 & $^{  +1.7 }_{  -1.9 }$ & [Ne~{\sc iii}]   & F1         & 2p4 3P     & 2p4 1D     &          3 &        5    \\
  3972.56 &   3970.07 &   11.2 & $\pm$   1.1 &  14.9 & $^{  +1.7 }_{  -2.0 }$ & H~{\sc i}        & H7         & 2p+ 2P*    & 7d+ 2D     &          8 &       98    \\
  4104.26 &   4101.74 &   23.0 & $\pm$   1.0 &  26.4 & $^{  +2.0 }_{  -2.2 }$ & H~{\sc i}        & H6         & 2p+ 2P*    & 6d+ 2D     &          8 &       72    \\
  4343.04 &   4340.47 &   43.2 & $\pm$   2.3 &  46.7 & $\pm$   2.1            & H~{\sc i}        & H5         & 2p+ 2P*    & 5d+ 2D     &          8 &       50    \\
  4365.79 &   4363.21 &    4.6 & $\pm$   0.7 &   6.3 & $^{  +0.8 }_{  -0.9 }$ & [O~{\sc iii}]    & F2         & 2p2 1D     & 2p2 1S     &          5 &        1    \\
  4474.40 &   4471.50 &    4.9 & $\pm$   0.5 &   5.7 & $^{  +0.6 }_{  -0.6 }$ & He~{\sc i}       & V14        & 2p 3P*     & 4d 3D      &          9 &       15    \\
  4544.54 &   4541.59 &    2.0 & $\pm$   0.6 &   2.9 & $^{  +0.6 }_{  -0.6 }$ & He~{\sc ii}      & 4.9        & 4f+ 2F*    & 9g+ 2G     &         32 &        *    \\
  4688.64 &   4685.68 &    8.3 & $\pm$   1.0 &   9.5 & $^{  +1.0 }_{  -1.1 }$ & He~{\sc ii}      & 3.4        & 3d+ 2D     & 4f+ 2F*    &         18 &       32    \\
  4864.28 &   4861.33 &  100.0 & $\pm$   4.4 & 100.0 & $\pm$   4.3            & H~{\sc i}        & H4         & 2p+ 2P*    & 4d+ 2D     &          8 &       32    \\
  4924.92 &   4921.93 &    2.3 & $\pm$   0.8 &   2.9 & $^{  +0.8 }_{  -0.8 }$ & He~{\sc i}       & V48        & 2p 1P*     & 4d 1D      &          3 &        5    \\
  4961.92 &   4958.91 &  208.0 & $\pm$   4.6 & 204.0 & $\pm$   5.0            & [O~{\sc iii}]    & F1         & 2p2 3P     & 2p2 1D     &          3 &        5    \\
  5009.88 &   5006.84 &  636.0 & $\pm$  12.9 & 613.0 & $\pm$  15.0            & [O~{\sc iii}]    & F1         & 2p2 3P     & 2p2 1D     &          5 &        5    \\
  5880.33 &   5875.66 &   12.6 & $\pm$   2.2 &  11.2 & $^{  +1.9 }_{  -2.3 }$ & He~{\sc i}       & V11        & 2p 3P*     & 3d 3D      &          9 &       15    \\
  6567.76 &   6562.77 &  349.3 & $\pm$  22.4 & 268.0 & $^{ +37.0 }_{ -43.0 }$ & H~{\sc i}        & H3         & 2p+ 2P*    & 3d+ 2D     &          8 &       18    \\
  7141.24 &   7135.80 &   19.0 & $\pm$   3.0 &  12.2 & $^{  +2.6 }_{  -3.3 }$ & [Ar~{\sc iii}]   & F1         & 3p4 3P     & 3p4 1D     &          5 &        5    \\
  7599.95 &   7592.74 &   15.2 & $\pm$   3.9 &  11.0 & $^{  +3.0 }_{  -3.6 }$ & He~{\sc ii}      & 5.1        & 5g+ 2G     & 10f+ 2H*   &         50 &        *    \\
 \hline
 \end{tabular}
 \end{table*}

\section{Nebular spectrum}
\label{sec:neatalfamoc}
\subsection{Empirical analysis}

We measured the emission line fluxes in the FORS2 spectrum of PN~G283.7$-$05.1 using {\sc alfa} (\citealt{wesson2016}), and calculated the physical conditions and ionic abundances using {\sc neat} (\citealt{wesson2012}). The shallowness of the observations (only about 15 emission lines are detected) means that the resulting information is sparse; no density diagnostics were available, and only the [O~{\sc iii}] line temperature diagnostic was available. {\sc neat} defaults to a density estimate of 1000\,cm$^{-3}$ in the absence of an observed value. Abundances could be estimated only for He, O, Ne, and Ar. For each of the heavy elements, only a single ionisation state was observed, and thus it was not possible to calculate  a reliable ionisation correction factor. The measured line fluxes are shown in Table~\ref{linelist}, and the calculations of the physical conditions are summarised in Table~\ref{empirical}.

\begin{table}
\caption{Empirical analysis of the FORS2 spectrum of PN~G283.7$-$05.1.  Abundance ratios are expressed by number.}
\label{empirical}
\centering
\begin{tabular}{ll}
\hline
Quantity          & Value \\
\hline
c(H$\beta$)       & 0.4$\pm$0.2 \\
Te([O~{\sc iii}]) & 11\,700$\pm$600\,K \\
Ne (assumed) & 1\,000\,cm$^{-3}$\\
\vspace{0.1cm}\\
He$^{+}$/H                          & ${  0.09\pm0.01}$ \\
He$^{2+}$/H                         & ${  7.95\times 10^{ -3}}^{+  9.10\times 10^{ -4}}_{ -8.10\times 10^{ -4}}$ \\
He/H                                & ${  0.10\pm0.01}$ \\
\vspace{0.1cm}\\
O$^{2+}$/H                          & ${  1.29\times 10^{ -4}}^{+  2.30\times 10^{ -5}}_{ -1.90\times 10^{ -5}}$ \\
Ne$^{2+}$/H                         & ${  3.15\times 10^{ -5}}^{+  5.70\times 10^{ -6}}_{ -4.90\times 10^{ -6}}$ \\
Ar$^{2+}$/H                         & ${  7.87\times 10^{ -7}}^{+  2.64\times 10^{ -7}}_{ -1.98\times 10^{ -7}}$ \\
\hline
\end{tabular}
\end{table}

\subsection{Photoionisation model}

Although the observed spectrum was shallow, it nevertheless allowed us to construct a simple photoionisation model to determine whether the nebular spectrum is consistent with the central star parameters derived in Section~\ref{sec:phoebe}. We matched our model to a subset of clearly detected emission lines: He~{\sc i} and He~{\sc ii} lines, which indicate the ionisation balance; [O~{\sc iii}] lines, which are strongly sensitive to the temperature; and bright Ne and Ar emission lines. The model is also constrained by the weakness of the [N~{\sc ii}] lines: the $\lambda$6584 line may be marginally detected, while $\lambda$6548 is not detected.

We used {\sc mocassin} v2.02.73 \citep{ercolano2003,ercolano2005} to calculate the model spectrum. We assumed a simple shell with a uniform density, with an outer radius of 2$\times$10$^{18}$cm (consistent with the observed angular size and distance determined in section \ref{sec:phoebe}). The ionising source was assumed to be a blackbody. We then varied the luminosity and temperature of the ionising source, starting from the values of 3275L$_\odot$ and 93kK derived in the previous section, and the density and elemental abundances of the gas. The He/H abundance was initially set to the empirically derived value of 0.096, while the abundances of heavy elements were initially set to the average PN values given in \citet{kingsburgh1994}. C and S lines are not observed, and are only indirectly constrained by the strong dependence of the nebular temperature on heavy element abundances.

In our initial set of models the [Ne~{\sc iii}] 3868 line was consistently overpredicted, while helium lines were underpredicted, and so we ran a revised set of models with Ne/O by number reduced from 0.25 to 0.17, and He/H increased to 0.12. To offset the resulting increase in nebular temperature, we increased the observationally unconstrained C/H and S/H abundances by a factor of 1.4.

After determining abundances that give a reasonable fit to all lines in the spectrum, we then ran a grid of 700 models in which we varied the hydrogen number density, the central star luminosity and temperature, and the abundances of heavy elements uniformly by factors of 0.75--1.25. The parameter space investigated is listed in Table~\ref{modeltable}.

In the final grid of 700 models, the formally best-fitting model (as calculated by summing $\chi^2$ values with all the fitted lines weighted equally) had a central star temperature of 100kK and luminosity of 3275L$_\odot$, but five models could provide fits to all the observed lines to within their observational uncertainties and several others provided close fits that would require only minor adjustments to bring into excellent agreement with the observations. Considering the 12 models for which all predicted line fluxes were within 2$\sigma$ of the observed values, this set had central star luminosities between 2570 and 3850L$_\odot$, temperatures between 93kK and 100kK, and hydrogen number densities of either 500 or 1000\,cm$^{-3}$.  The best-fitting model using the exact central star parameters derived in Section~\ref{sec:phoebe} predicted all the lines to within their uncertainties except He~{\sc ii} $\lambda$4686, which was underpredicted. Given the simplicity of the model, in particular the assumption of a blackbody, when He~{\sc ii} line fluxes are very sensitive to the shape of the ionising spectrum, this discrepancy is not of great significance. This modelling demonstrates that the nebular spectrum is consistent with the central star parameters derived above.

The non-detection of [N~{\sc ii}] lines in the observed spectrum results in an extremely low N/H abundances in the models, with values around 2$\times$10$^{-6}$ required. This is a factor of 100 less than the average PN value from \citet{kingsburgh1994}. However, the fraction of N in ionised nebulae in the form of N$^{+}$ is usually very small and will be very sensitive to the shape of the ionising spectrum. Much deeper optical spectra providing sufficient constraints to construct a more detailed photoionisation model would be required to confirm whether this apparently extremely low nitrogen abundance is real.

Table~\ref{modellines} lists the observed and predicted line fluxes from both the formally best-fitting model, and from the best-fitting model using the luminosity and temperature derived in Section~\ref{sec:phoebe}.

\begin{table}
\caption{Parameter space covered by our photoionisation models and the best-fitting values. Abundance ratios are expressed by number.}
\label{modeltable}
\begin{tabular}{lll}
\hline
Parameter & Range investigated & Best-fitting value \\
\hline
T$_*$ (kK)  & 80, 90, 93, 96, 100, 110, 120 & 100 \\
L$_*$ (L$_\odot$) & 1285, 2570, 3275, 3850, 5140 & 3275 \\
n$_H$ (cm$^{-3}$) & 500, 1000, 2000, 4000 & 1000 \\
He/H & 0.096, 0.12 & 0.12 \\

C/H & 5.77 - 9.62$\times$10$^{-4}$ & 8.47$\times$10$^{-4}$ \\
N/H & 1.50 - 2.50$\times$10$^{-6}$ & 2.20$\times$10$^{-6}$ \\
O/H & 1.46 - 2.44$\times$10$^{-4}$ & 2.15$\times$10$^{-4}$ \\
Ne/H & 2.48 - 4.12$\times$10$^{-5}$ & 3.63$\times$10$^{-5}$ \\
S/H & 0.87 - 1.45$\times$10$^{-6}$ & 1.28$\times$10$^{-5}$ \\
Ar/H & 0.90 - 1.50$\times$10$^{-6}$ & 1.32$\times$10$^{-6}$ \\

\hline
\end{tabular}
\end{table}

\begin{table}
    \caption{Predicted emission line fluxes (H$\beta$=100) from the model providing the best fit overall (M1), and the best-fitting model with T$_\mathrm{eff}$ and L from Section 3 (M2).}
    \label{modellines}
    \begin{tabular}{lllll}
    \hline
Wavelength (\AA) & Ion & Observed flux & M1 & M2 \\
\hline
3868 & [Ne~{\sc iii}] &  53.5 $\pm$  6.0 &  55.4 & 50.0\\
4363 & [O~{\sc iii}]  &   6.3 $\pm$  0.8 &   6.2 &  6.6\\
4471 & He~{\sc i}     &   5.7 $\pm$  0.6 &   6.2 &  6.1\\
4686 & He~{\sc ii}    &   9.5 $\pm$  1.0 &   9.9 &  8.0\\
4959 & [O~{\sc iii}]  & 204.0 $\pm$  5.0 & 205.3 & 203.9\\
5007 & [O~{\sc iii}]  & 613.0 $\pm$ 15.0 & 612.6 & 608.4\\
5876 & He~{\sc i}     &  11.2 $\pm$  5.0 &  16.9 & 16.6\\
6548 & [N~{\sc ii}]    &  $<$8            &   2.5 &  1.6\\
6584 & [N~{\sc ii}]   &  $<$10           &   7.7 &  5.1\\
7135 & [Ar~{\sc iii}] &  12.2 $\pm$  3.0 &  12.9 & 10.1\\
\hline
    \end{tabular}
\end{table}

\section{Discussion}
\label{sec:disc}

We have presented the discovery and characterisation of the post-CE binary nucleus of \png{}.  Simultaneous light and radial velocity curve modelling with the \textsc{phoebe}2 code reveal the system to comprise a highly irradiated M-type main-sequence star in a 5.9-hour orbit with the nebular progenitor.

The RV curve of the secondary component was derived using a complex of irradiated emission lines.  The treatment of such irradiated atmospheres is challenging, with almost all modelling efforts resorting to a bolometric treatment where such emission lines are ignored \citep[see][for a more detailed discussion]{horvat19}.  There is some uncertainty as to what region of the secondary they are representative of.  Assuming that they are representative of the stellar photocentre (whose the location  is dominated by the high levels of irradiation), the primary's mass is found to be 0.34$\pm$0.05~M$_\odot$,  inconsistent with a post-AGB core mass.  A maximum primary mass was then derived by adjusting the modelling to assume that the irradiated emission lines originate only from the very innermost point on the secondary's irradiated hemisphere, leading to a primary mass determination of 0.42$\pm$0.05~M$_\odot$,  spanning acceptable post-AGB and post-RGB core mass.
In both cases, the secondary radius and temperature are found to be slightly higher than those found for field stars of the same mass \citep[as is generally the case in young post-CE binaries;][]{jones15}.  Similarly, the model primary mass, temperature, and radius are found to be inconsistent with evolutionary tracks \citep{hall13,MillerBertolami2016}. Despite the complexities of evolving through a CE and the uncertainties involved in our modelling, it is difficult to envisage how a CE interaction could lead to configurations that may simply be too difficult to obtain as a solution to the stellar structure equations. In particular, a central star mass as low as the one derived here is difficult to reconcile with AGB evolution, even if it were cut short by a CE.

A sanity check on the model temperatures and radii was performed by calculating synthetic reddened magnitudes for the bands in which the binary was observed, implying a distance of roughly 9~kpc (borderline consistent with the Gaia parallax).  At this distance, the PN measures roughly 1~pc and lies between the Sagittarius and Perseus arms of the Galaxy.  A further check was performed through comparison of observed nebular emission-line fluxes to those from a simple photoionisation model, the physical parameters of which were consistent with those derived from the simultaneous light and RV curve modelling.

While there is significant uncertainty, the results collectively indicate that the central star mass of \png{} is roughly 0.3--0.4~M$_\odot$, consistent with having experienced the common-envelope event while still on the red giant branch.
Only a handful of other post-CE binary central stars of PNe have been identified as possible post-RGB systems, the strongest of which being ESO~330-9 \citep{hillwig17}\footnote{The central star of the Stingray nebula was also identified by \citet{reindl14} as a possible post-RGB PN, but its central star has since evolved back towards the AGB consistent with having experienced a late thermal pulse rather than a CE evolution \citep{reindl17}.}.  Meanwhile, recent population synthesis models, based on the observed frequency of Sequence E variables, indicate that post-RGB PNe should be roughly as common as those produced by CE events during the AGB \citep{nie12}.  Similarly, the strong post-RGB peak at M$<0.4$~M$_\odot$ observed in the mass distribution of white dwarf main-sequence binaries discovered by SDSS is another strong indication that CE events on the RGB are both frequent and survivable \citep{rebassa-mansergas11}.

The discovery and characterisation of other post-RGB PNe and their central stars will be critical in constraining the importance and nature of this pathway.  They may even hold the key to elucidating other long-standing issues, such as the large abundance discrepancy factors observed in some post-CE PNe that have been speculatively linked to RGB CEs \citep{jones16}.

\begin{acknowledgements}
The authors thank the referee, Orsola De Marco, for her thorough and insightful report.

DJ, RLMC, JG-R and PR-G acknowledge support from the State Research Agency (AEI) of the Spanish Ministry of Science, Innovation and Universities (MCIU) and the European Regional Development Fund (FEDER) under grant AYA2017-83383-P.  DJ, and JG-R also acknowledge support under grant P/308614 financed by funds transferred from the Spanish Ministry of Science, Innovation and Universities, charged to the General State Budgets and with funds transferred from the General Budgets of the Autonomous Community of the Canary Islands by the Ministry of Economy, Industry, Trade and Knowledge. PS thanks the Polish National Center for Science (NCN) for support through grant
2015/18/A/ST9/00578.  JM acknowledges the support of the ERASMUS+ programme in the form of a traineeship grant.

The authors thankfully acknowledge the technical expertise and assistance provided by the Spanish Supercomputing Network (Red Espa\~nola de Supercomputaci\'on), as well as the computer resources used: the LaPalma Supercomputer, located at the Instituto de Astrof\'isica de Canarias.

This paper is based on observations made with ESO Telescopes at the La Silla Paranal Observatory under programme IDs 088.D-0573,  090.D-0449, 090.D-0693, 093.D-0038, 094.D-0091, 096.D-0234, 096.D-0237, 096.D-0800, 097.D-0351. This research has made use of the VizieR catalogue access tool, CDS, Strasbourg, France; the SIMBAD database, operated at CDS, Strasbourg, France; NASA's Astrophysics Data System; APLpy, an open-source plotting package for Python hosted at http://aplpy.github.com; Astropy,\footnote{http://www.astropy.org} a community-developed core Python package for Astronomy \citep{astropy:2013, astropy:2018}.

\end{acknowledgements}

\bibliographystyle{aa}
\bibliography{literature}

\end{document}